\documentclass[aps,prd,a4paper,onecolumn,amsmath,showpacs,superscriptaddress,nofootinbib,preprintnumbers,notitlepage]{revtex4-1}
 
\usepackage{verbatim}
\usepackage[T1]{fontenc}
\usepackage[utf8]{inputenc}
\usepackage[american]{babel}
\usepackage{graphicx,subcaption,caption}
\usepackage{epsfig}
\usepackage{graphicx,subcaption,caption}
\usepackage{pgfplots}
\usepackage{booktabs}
\usepackage{multirow}
\usepackage{dcolumn}
\usepackage{amsmath}
\usepackage{mathtools}
\usepackage{amsfonts}
\usepackage{amssymb}
\usepackage{epstopdf}
\usepackage{bm}
\usepackage{siunitx}
\usepackage{braket}
\usepackage{enumitem}
\usepackage{soul}
\usepackage{amsmath}

\DeclareMathOperator{\csch}{csch}

\DeclareMathOperator{\arctanh}{arctanh}

\usepackage{transparent}
\usepackage{pifont}

\definecolor{navyblue}{rgb}{0.0, 0.0, 0.5}
\definecolor{royalblue}{rgb}{0.25, 0.41, 0.88}
\definecolor{cadmiumgreen}{rgb}{0.0, 0.42, 0.24}
\definecolor{blue-violet}{rgb}{0.54, 0.17, 0.89}
\definecolor{darkviolet}{rgb}{0.58, 0.0, 0.83}
\definecolor{orange(colorwheel)}{rgb}{1.0, 0.5, 0.0}

\usepackage{hyperref}
\hypersetup{
    colorlinks=true, 
    linkcolor=royalblue, 
    citecolor=magenta}

\newcommand\be{\begin{equation}}
\newcommand\ee{\end{equation}}

\newcommand\bea{\begin{eqnarray}}
\newcommand\eea{\end{eqnarray}}

\usepackage{booktabs}
\usepackage{multirow}
\usepackage{dcolumn}
\usepackage{colortbl}

\definecolor{magenta(process)}{rgb}{1.0, 0.0, 0.56}

\definecolor{darkspringgreen}{rgb}{0.09, 0.45, 0.27}

\definecolor{royalblue(web)}{rgb}{0.25, 0.41, 0.88}

\begin{document}

\title{Quintessential constant-roll inflation}

\author{Mehdi Shokri}
\email{mehdishokriphysics@gmail.com}
\affiliation{School of Physics, Damghan University, P. O. Box 3671641167, Damghan, Iran}

\author{Jafar Sadeghi}
\email{pouriya@ipm.ir}
\affiliation{Department of Physics, University of Mazandaran, P. O. Box 47416-95447, Babolsar, Iran}
\affiliation{School of Physics, Damghan University, P. O. Box 3671641167, Damghan, Iran}

\author{Saeed Noori Gashti}
\email{saeed.noorigashti@stu.umz.ac.ir}
\affiliation{Department of Physics, University of Mazandaran, P. O. Box 47416-95447, Babolsar, Iran}

\preprint{}
\begin{abstract}
We investigate a single field model in the context of the constant-roll inflation in which inflaton moves down to the minimum point of the potential with a constant rate of rolling. We use a quintessential inflationary model obtained by a Lorentzian function which is dependent on the number of e-folds. We present the inflationary analysis for the model and find the observational constraints on the parameters space using the observations of CMB anisotropies \textit{i.e.} the Planck and Keck/array datasets. We find the observationally acceptable values of the Width of the Lorentzian function as $0.3<\Gamma\leq0.5$ at the $68\%$ CL and $\Gamma\leq0.3$ at the $95\%$ CL when $\xi=120$, $|\beta|=0.02$ and $N=60$. Also, we acquire the observationally favoured values of the amplitude of the Lorentzian function as $400<\xi\leq600$ at the $68\%$ CL and $\xi\leq400$ at the $95\%$ CL when $\Gamma=0.1$, $|\beta|=0.02$ and $N=60$. Moreover, we study the model from the Weak Gravity Conjecture approach using the swampland criteria.  
\end{abstract}
\maketitle
\section{Introduction}
 The theory of inflation presents the most successful model to explain the phenomena in the early universe by considering a rapid and large expansion in the scale factor. In addition to removing the shortcomings of the hot big bang model, inflation provides an acceptable mechanism to generate the primordial density perturbations which are known as the main candidate for the structure formation of the universe and also the temperature anisotropy of the cosmic background photons. Moreover, the tensor perturbations which form the primordial gravitational waves have been produced during the inflationary epoch \cite{Guth,Kazanas:1980tx,Linde:1981my,Albrecht:1982wi,Lyth:1998xn}. In the standard picture, a single scalar field is the main responsible to push the inflationary era. Inflaton slowly rolls down from the peak of the potential to the down under the slow-roll approximation. Finally, the inflaton decays to the particles and releases enough energy to reheat the supercooled universe \cite{Kofman2,Shtanov}. Using the inflationary observations coming from the Planck and other related satellites, we find a few single field models in good agreement with the observational datasets \cite{staro,barrow,kallosh1} and the majority of them are ruled out or restricted \cite{martin}. Also, these models do not show non-Gaussianity in their spectrum because of the existence of the uncorrelated modes. \cite{Chen}. Therefore, if the future observations show a non-Gaussianity feature in the spectrum, the single field models will be situated in uncertainty. To escape such problems, it is proposed to go beyond the slow-roll approximation in order to create some non-Gaussianity in the perturbations spectrum of single field models. Hence, constant-roll inflation is suggested in which the inflaton rolls down with a constant rate
\begin{equation}
\ddot{\varphi}=\beta H\dot{\varphi}\,,
\label{1}
\end{equation}
where $\beta=-(3+\alpha)$ and $\alpha$ is a non-zero parameter \cite{martin2,Motohashi1,Motohashi2}. Going beyond the slow-roll approach is also addressed in the \textit{ultra slow-roll} inflation in which we deal with a non-negligible $\ddot{\varphi}$ in the Klein-Gordon equation $\ddot{\varphi}=3H\dot{\varphi}$ \cite{Inoue,Pattison,Kinney,Namjoo}. It is also found in the \textit{fast-roll} realm where a fast-rolling phase is assumed at the start of inflation and then is followed by the standard slow-roll after a few e-folds \cite{Contaldi,Lello,Hazra}. Recently, the constant-roll inflationary approach has been intensively engaged so that one can find a wide range of the inflationary models in this content, see, \textit{e.g.}, \cite{Odintsov,Nojiri,Sebastiani,Motohashi9,Cicciarella,Anguelova,Ito,karam1,Ghersi,Lin,Micu,Oliveros,Motohashi3,oi1,neee,oi2,Kamali,diego,sh1,sh2,oki,sh5,shokriii}. 

In this paper, we consider a quintessential inflationary model \cite{vil} introduced by a Lorentzian function which is dependent on the number of e-folds in order to obtain the slow-roll parameters. This approach studied in \cite{benn1,ben1,ben2,benn2} for a single field model. It was shown that the corresponding inflationary model in addition to explain the early time acceleration can describe the late-time acceleration phase (dark energy), asymptotically. The above discussion motivates us to arrange the paper as follows. In \S II, we present the constant-roll inflationary analysis for the model using the Lorentzian function. We analyze the obtained results of the model by comparison with the inflationary observations in \S III. In \S IV, we investigate the model from viewpoint of the Weak Gravity Conjecture using the swampland criteria. Conclusions are presented in \S V.
\section{Constant-roll Inflation}
We start with a Lorentzian function for the first slow-roll parameter \cite{ben1}
\begin{equation}
\epsilon_{1}\equiv\epsilon(N)=\frac{2\xi\Gamma}{\pi(4N^{2}+\Gamma^{2})},
\label{2}
\end{equation}
as a function of e-folds $N$ which is defined by $e^{N}=(a_{f}/a_{i})$. Also, $\xi$ and $\Gamma$ are the amplitude and the width of the Lorentzian function. Using the definition of the first slow-roll parameter $\epsilon_{1}=-H'/H$, we find an expression for the Hubble parameter 
\begin{equation}
H=C\exp{\Big(-\frac{\xi}{\pi}\arctan{(\frac{2N}{\Gamma})}\Big)}  
\label{3},
\end{equation}
where $C$ is an integration constant. We can calculate other slow-roll parameters using the generator formula $\epsilon_{n+1}=d\ln|\epsilon_{n}|/dN$ as follow
\begin{equation}
\epsilon_{2}=-\frac{8N}{\Gamma^{2}+4N^{2}},\hspace{1cm}\epsilon_{3}=\frac{1}{N}-\frac{8N}{\Gamma^{2}+4N^{2}}.
\label{4}
\end{equation}
In order to study the model in the context of the constant-roll approach, we set 
\begin{equation}
\epsilon_{2}=-\frac{\ddot{\varphi}}{H\dot{\varphi}}\equiv=-\beta,
\label{5}
\end{equation}
that leads to 
\begin{equation}
\beta=\frac{8N}{\Gamma^{2}+4N^{2}}.
\label{6}
\end{equation}
Now, let's apply the formalism for a single field inflationary model. We consider the action
\begin{equation}
S=\int{d^{4}x\sqrt{-g}\bigg(R-\frac{1}{2}g^{\mu\nu}\partial_{\mu}\varphi\partial_{\nu}\varphi-V(\varphi)\bigg)}
\label{7}
\end{equation}
where $g$ is the determinant of the metric $g_{\mu\nu}$, $R=g^{\mu\nu}R_{\mu\nu}$ is the Ricci scalar. Also, here we suppose $\kappa^{2}\equiv8\pi G=1$. Using the energy-momentum tensor of a perfect fluid and Friedman-Robertson-Walker (FRW) metric $ds^{2}=-dt^{2}+a(t)^{2}(dx^{2}+dy^{2}+dz^{2})$ for a  homogeneous, isotropic and spatially flat universe, the dynamical equations are given by
\begin{equation}
3H^{2}=\frac{\dot{\varphi}^{2}}{2}+V,\hspace{1cm}2\dot{H}=-\dot{\varphi}^{2},
\label{8}
\end{equation}
where $H\equiv\dot{a}/a$ is the Hubble parameter, and the dot represents the derivative with respect to $\varphi$. Moreover, by varying the action (\ref{7}) with respect to  $\varphi$, we find the Klein-Gordon equation 
\begin{equation}
\ddot{\varphi}+3H\dot{\varphi}+\frac{d V}{d\varphi}=0.
\label{9}
\end{equation}
From the Friedmann equation (\ref{8}) and using the obtained expression of the Hubble parameter (\ref{3}) and the relation (\ref{6}), we find
\begin{equation}
V(N)=Ce^{-\frac{2\xi}{\pi}\arctanh{\big(\frac{2N}{\Gamma}\big)}}\bigg(1-\frac{\Gamma\xi\beta}{12\pi N}\bigg), 
\label{10}
\end{equation}
\begin{figure*}[!hbtp]
	\centering
	\includegraphics[width=.32\textwidth,keepaspectratio]{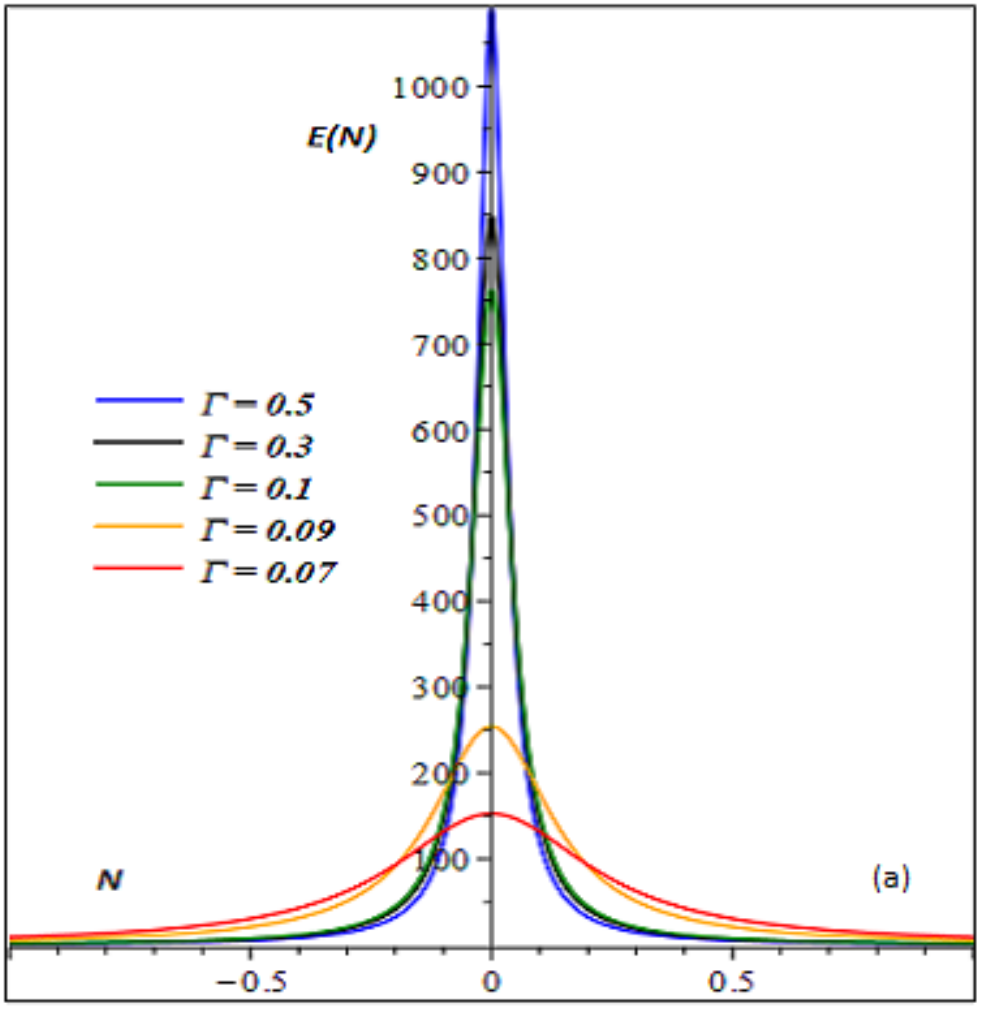}
	\includegraphics[width=.32\textwidth,keepaspectratio]{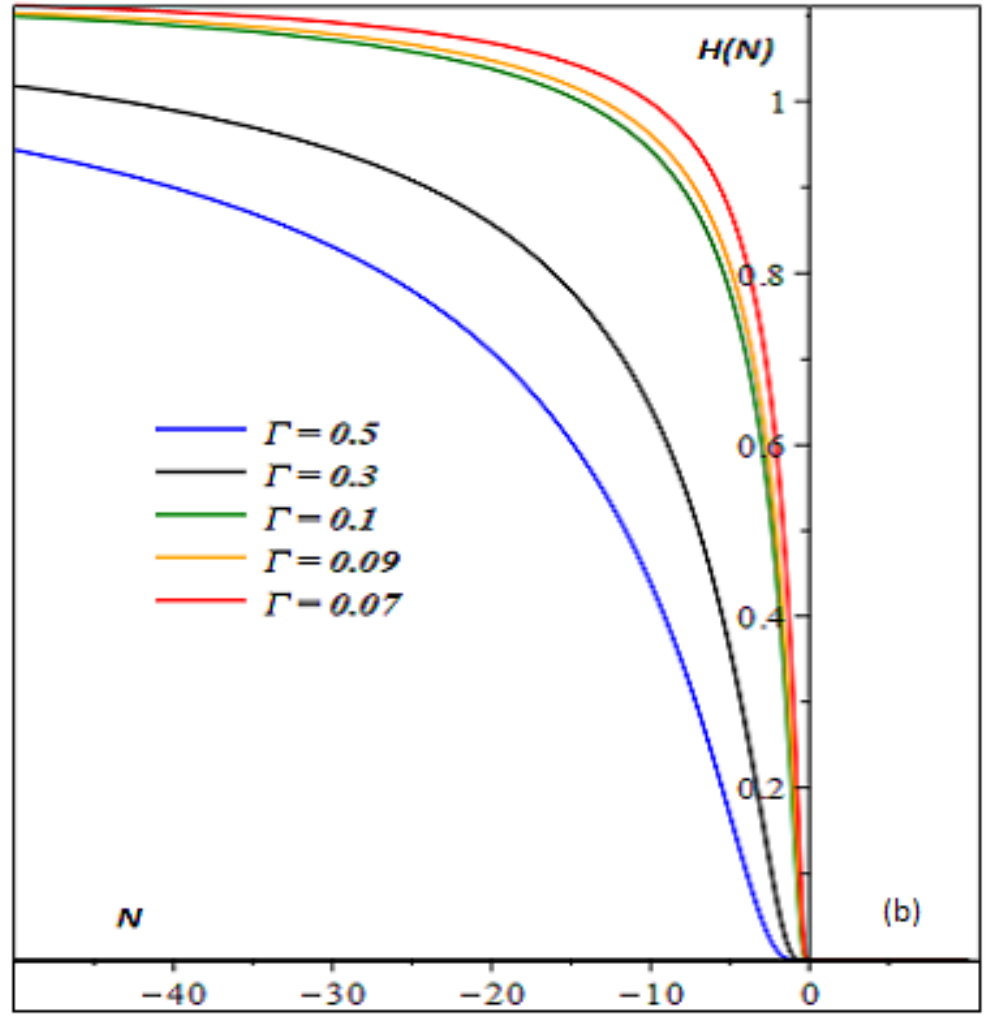}
	\includegraphics[width=.32\textwidth,keepaspectratio]{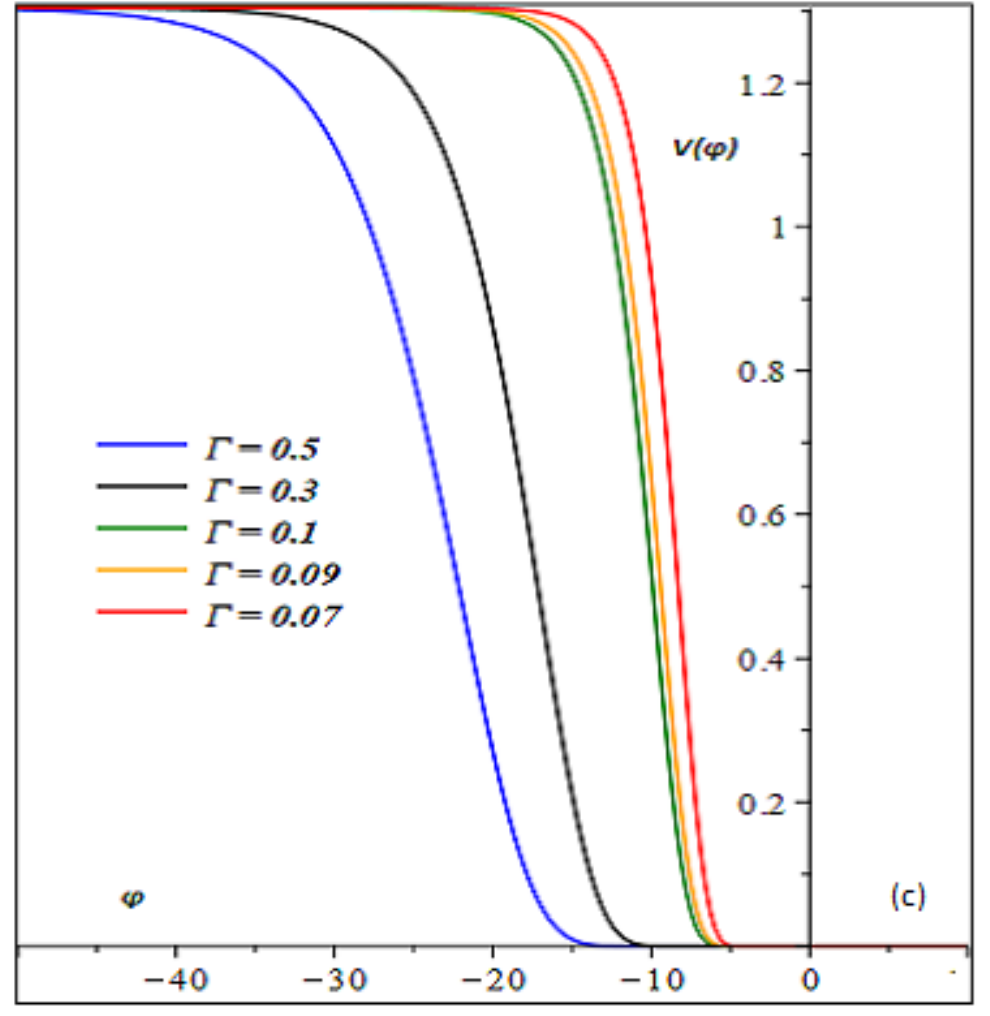}
	\caption{The first slow-roll parameter $\epsilon$ (\ref{2}) and the Hubble parameter $H$ (\ref{3}) plotted versus the number of e-folds $N$ (a and b). The potential (\ref{12}) plotted versus the scalar field $\varphi$ (c). The plots are drawn for different values of the width $\Gamma$ when $|\beta|=0.02$ and the amplitude $\xi$ is $\sim120$. Also, $C=10^{-29}$ for the panel (b) and $C=10^{-52}$ for the panel (c).}
	\label{fig1}
\end{figure*}
that using the number of e-folds 
\begin{equation}
N=\frac{\Gamma}{2}\sinh{\Big(\sqrt{\frac{\pi}{\xi\Gamma}}\varphi\Big)},   
\label{11}
\end{equation}
the potential (\ref{10}) takes the following form
\begin{equation}
V(\varphi)=Ce^{-\frac{2\xi}{\pi}\arctanh{\Big(\sinh{\big(\sqrt{\frac{\pi}{\xi\Gamma}}\varphi\big)}\Big)}}\bigg(1-\frac{\xi\beta}{6\pi}\csch\big(\sqrt{\frac{\pi}{\xi\Gamma}}\varphi\big)\bigg),
\label{12}
\end{equation}
where $C$ is the integration constant. The panels (a) and (b) of Figure \ref{fig1} show the behaviour of the first slow-roll parameter $\epsilon$ (\ref{2}) and the Hubble parameter $H$ (\ref{3}) versus the number of e-folds $N$. The panel (c) presents the behaviour of the potential (\ref{12}) versus the scalar field $\varphi$. The panels are drawn for different values of the width $\Gamma$ when $|\beta|=0.02$ and the amplitude $\xi\sim120$. Also, $C=10^{-29}$ for the panel (b) and  $C=10^{-52}$ for the panel (c). In the presence of the constant-roll parameter $\beta$, the potential still shows a quintessential inflationary manner since it supplies the late-time acceleration condition, asymptotically. The slow-roll parameters of the model
\begin{equation}
\epsilon\equiv\frac{1}{2}\bigg(\frac{V'}{V}\bigg)^{2},\hspace{1cm}\eta\equiv\frac{V''}{V},\hspace{1cm}\zeta^{2}\equiv\frac{V'V'''}{V^{2}},
\end{equation}
are calculated as shown in (\ref{a1} - \ref{a3}). Note that inflation ends when the condition $\epsilon = 1$ or $\eta = 1$ is fulfilled. The number of e-folds of the model takes the following form 
\begin{equation}
N\equiv\int^{\varphi_{i}}_{\varphi_{f}}{\frac{1}{\sqrt{2\epsilon}}}d\phi\simeq\frac{3\Gamma}{(\beta-12)}e^{\sqrt{\frac{\pi}{\xi\Gamma}}\varphi},
\label{14}
\end{equation}
and then the spectral index and tensor-to-scalar ratio
\begin{equation}
n_{s}=1-6\epsilon+2\eta,\quad\quad\quad r=16\epsilon,
\label{17}  
\end{equation}
are obtained as shown in (\ref{a4}) and (\ref{a5}).
\section{Constraints from CMB Anisotropies}
\begin{figure*}[!hbtp]
	\centering
	\includegraphics[width=.47\textwidth,keepaspectratio]{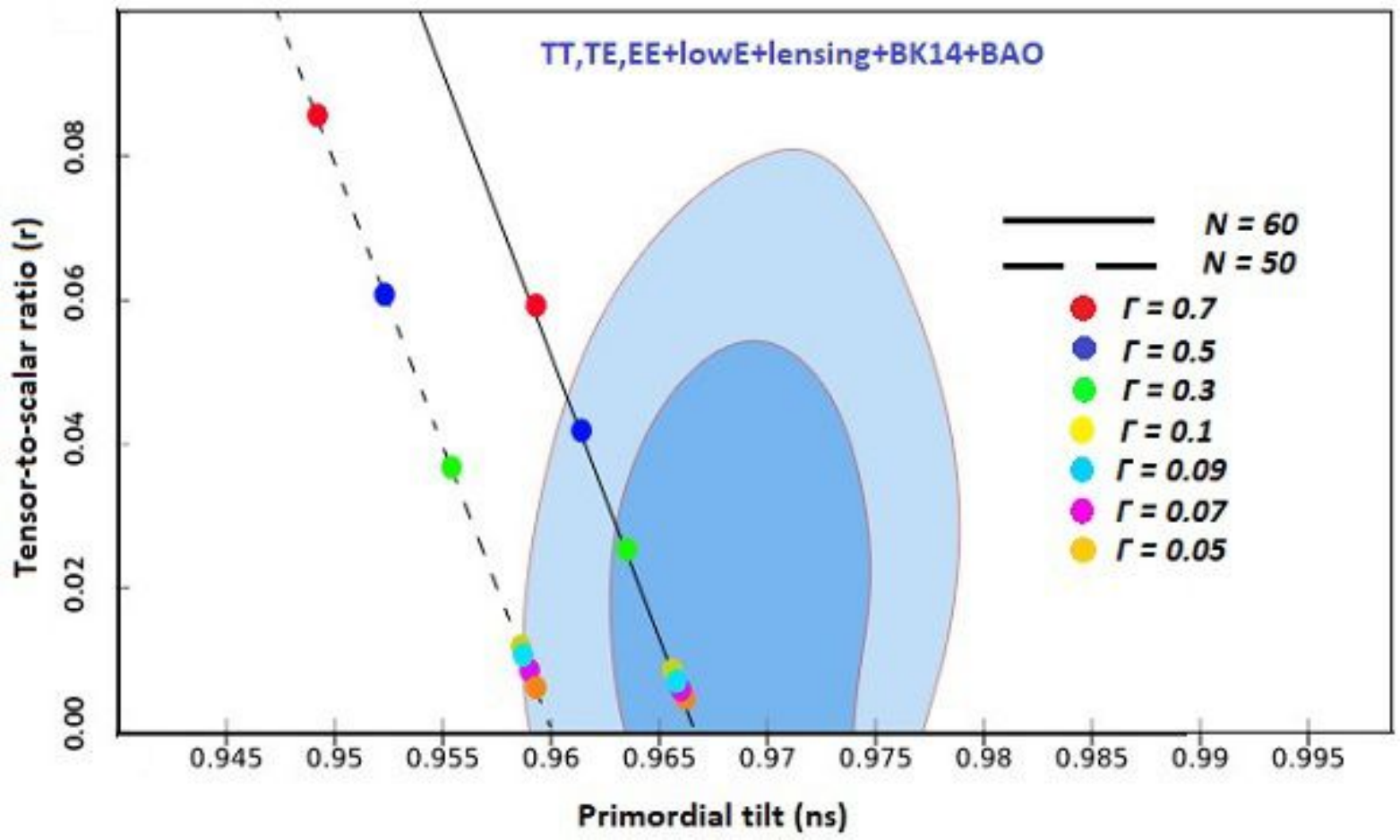}
	\includegraphics[width=.465\textwidth,keepaspectratio]{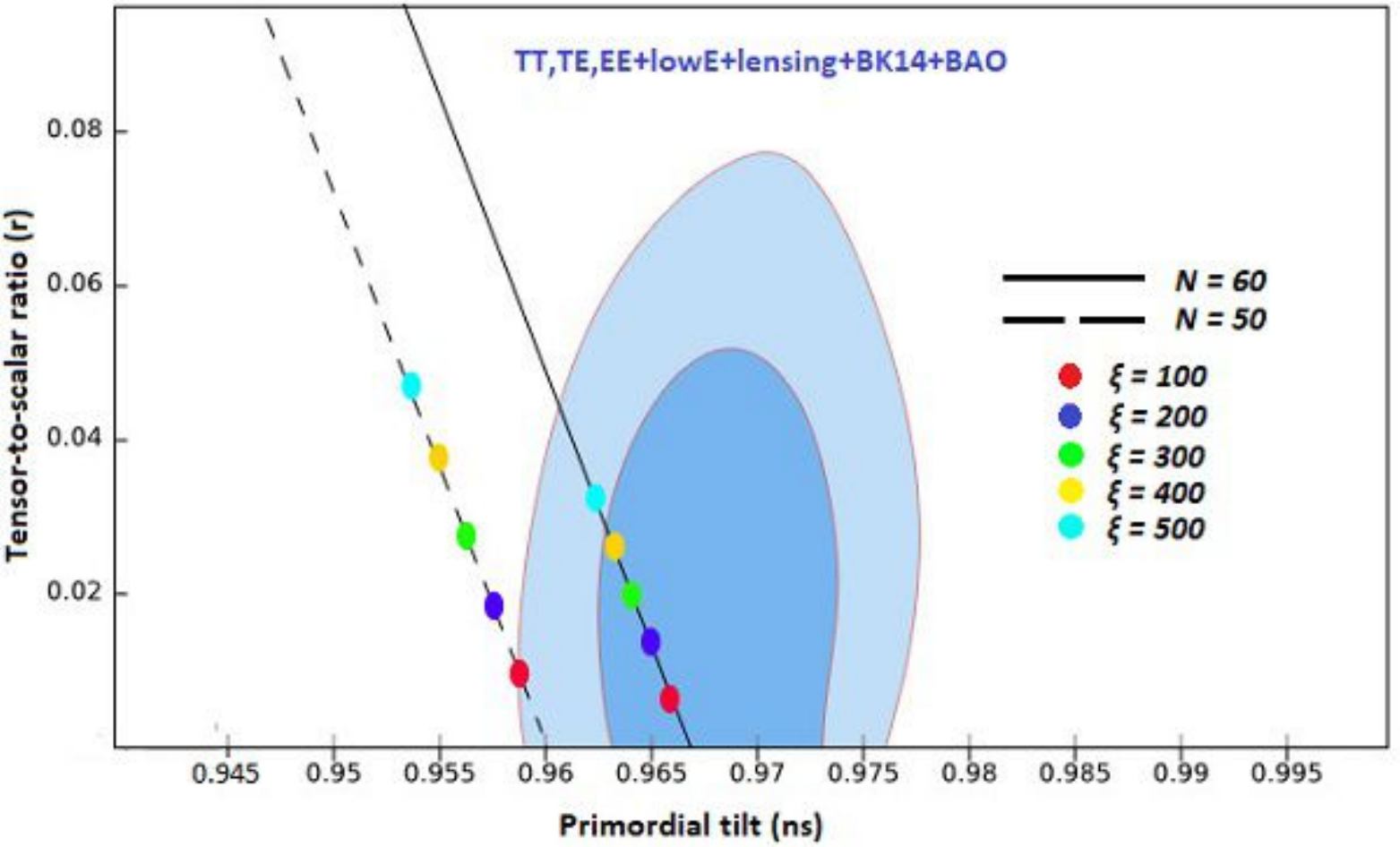}
	\caption{The marginalized joint 68\% and 95\% CL regions for $n_{s}$ and $r$ at $k = 0.002$ Mpc$^{-1}$ from Planck in combination with BK14+BAO data \cite{cmb} and the $n_{s}-r$ constraints on the model (\ref{12}). The dashed and solid lines represent $N=50$ and $N=60$, respectively. The left panel is plotted for different values of the width $\Gamma$ when $|\beta|=0.02$ and the amplitude $\xi$ is $\sim120$. The right panel is plotted for different values of the amplitude $\xi$ when $|\beta|=0.02$ and the width $\Gamma$ is $0.1$.}
	\label{fig2}
\end{figure*}

In Figure \ref{fig2}, we present the $n_{s} - r$ constraints coming from the marginalized joint 68\% and 95\% CL regions of the Planck 2018 in combination with BK14+BAO data on the Lorentzian quintessential inflationary model (\ref{12}) introduced in the context of the constant-roll approach. The dashed and solid lines represent $N=50$ and $N=60$, respectively.

The left panel of Figure \ref{2} is plotted for different values of the width $\Gamma$ when $|\beta|=0.02$ and the amplitude $\xi$ is $\sim120$. For $N=50$, the majority of the points show a reasonable value of tensor-to-scalar ratio $r$ despite disfavored values of the spectral index $n_{s}$.
\begin{figure*}[!hbtp]
	\centering
	\includegraphics[width=.35\textwidth,keepaspectratio]{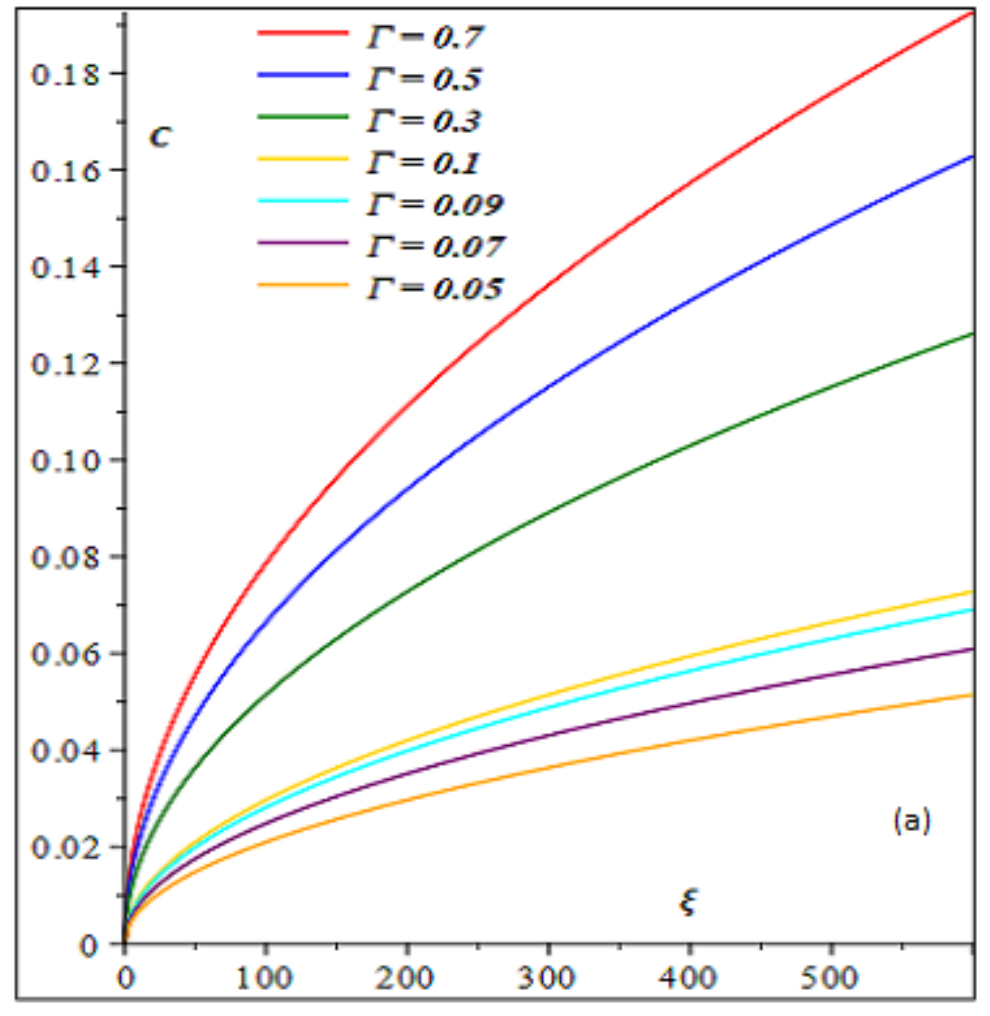}
	\hspace{0.5cm}
	\includegraphics[width=.35\textwidth,keepaspectratio]{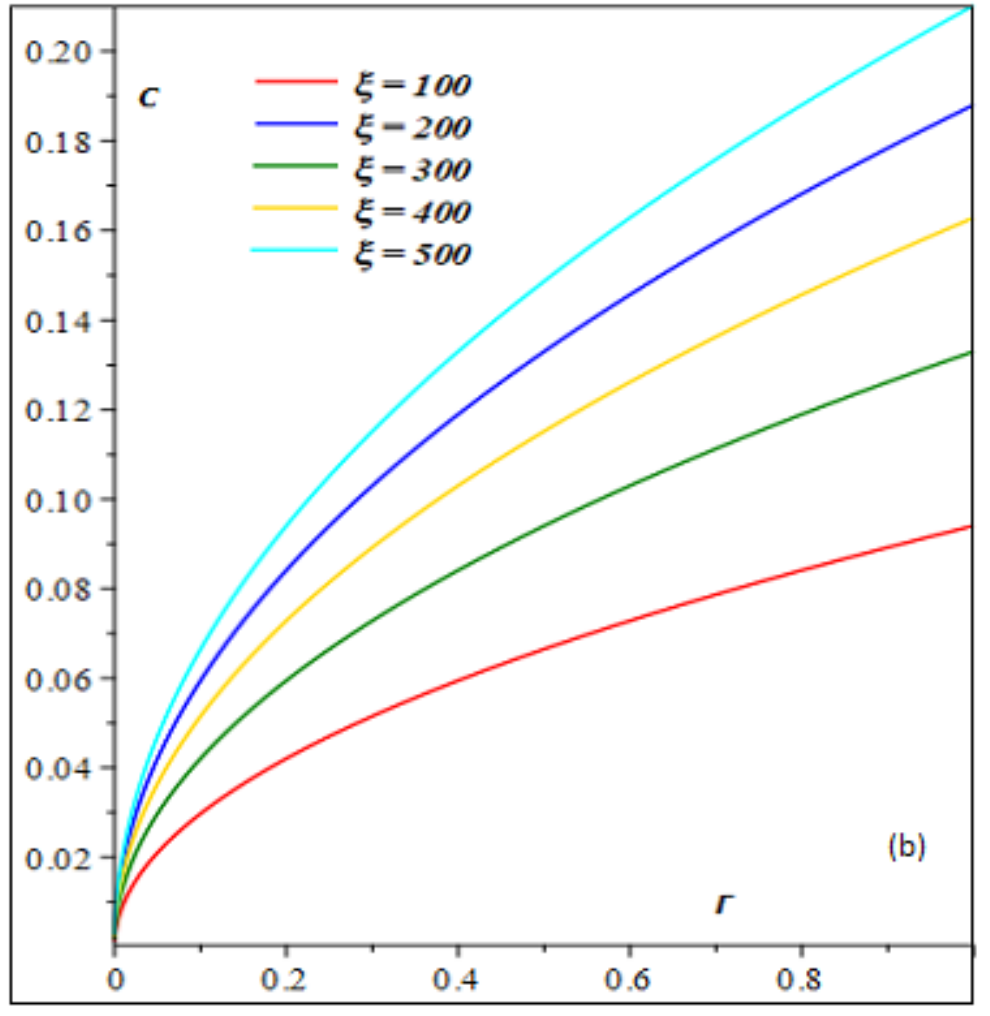}
	\end{figure*}
\vspace{2cm}
\begin{figure*}[!hbtp]
	\includegraphics[width=.35\textwidth,keepaspectratio]{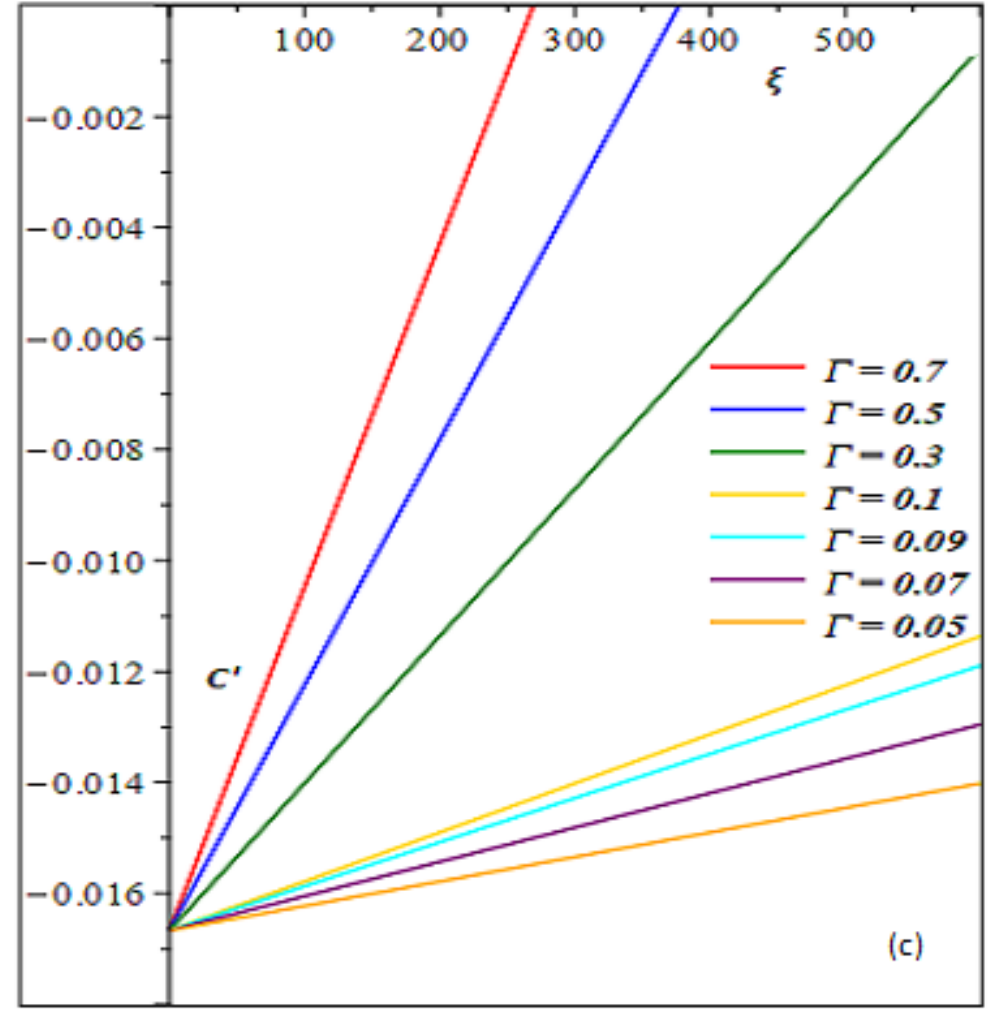}
	\hspace{0.5cm}
	\includegraphics[width=.35\textwidth,keepaspectratio]{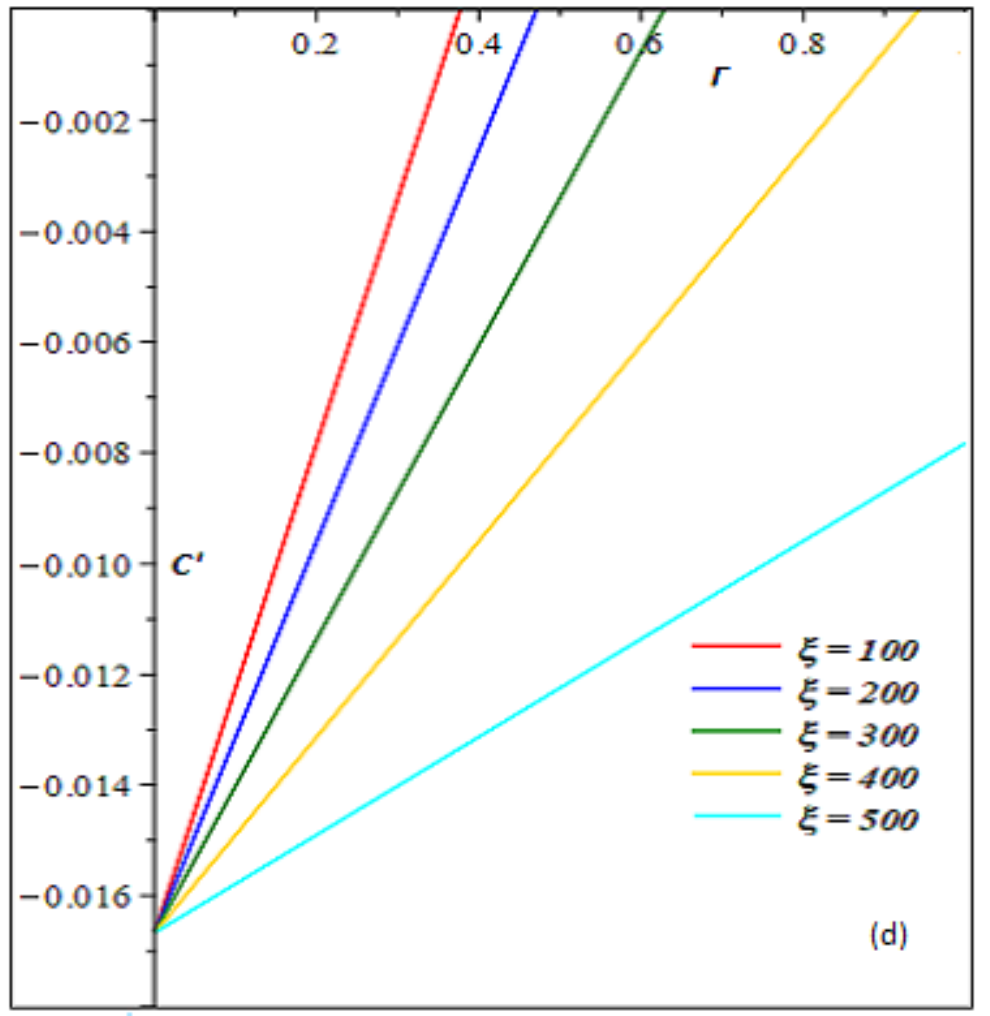}
	\caption{The behaviour of the swampland parameter $c$ versus the amplitude $\xi$ for different widths $\Gamma$ (a). The behaviour of the swampland parameter $c$ versus the width $\Gamma$ for different amplitudes $\xi$ (b). The behaviour of the swampland parameter $c'$ versus the amplitude $\xi$ for different widths $\Gamma$ (c). The behaviour of the swampland parameter $c'$ versus the width $\Gamma$ for different amplitudes $\xi$ (d). The plots are drawn for $|\beta|=0.02$ and $N=60$.}
	\label{fig3}
\end{figure*}
Considering the $68\%$ CL, we find that the obtained values of $r$ and $n_{s}$ for the width $\Gamma\leq0.1$ are in good agreement with the observations. Note that at the $95\%$, the panel does not show any compatibility with the observational datasets. For $N=60$, the situation is quite better than $N=50$. The panel reveals that the width $\Gamma>0.5$ is ruled out by the Planck data while the widths $0.3<\Gamma\leq0.5$ and $\Gamma\leq0.3$ show observationally acceptable values of $n_{s}$ and $r$ at the $68\%$ and $95\%$ CL, respectively. The right panel of Figure \ref{2} is plotted for different values of the amplitude $\xi$ when $|\beta|=0.02$ and the width $\Gamma$ is $0.1$. In the case of $N=50$, the obtained values of $r$ and $n_{s}$ are out of the observational region in exception the amplitude $\xi\leq100$ that is compatible with the observations at the $68\%$ CL. For $N=60$, the panel shows that the results are fully consistent with the CMB observations for $400<\xi\leq600$ and $\xi\leq400$ at the $68\%$ and $95\%$ CL, respectively.
\section{Swampland Criteria}
String theory as a developed theory in particle physics which can describe many cosmological phenomena in the early universe without taking into account a quantum gravity theory. String landscape and swampland regions included with the low-energy effective field theories are compatible and non-compatible with string theory, respectively. Hence, in order to embed the effective field theories into stringy quantum gravity, we require to distinguish two regions from each other. Weak Gravity Conjecture (WGC) is the most well-defined tool to achieve the demanded separation \cite{Vafa,Brennan,Ooguri,Palti,  Palti2,Obied}. Let's study the model from the viewpoint of WGC using the conditions of the swampland de sitter conjecture  
\begin{equation}
\sqrt{2\epsilon}\geq c,\hspace{1cm}|\eta|\leq -c',
\label{16}
\end{equation}
where $c$ and $c'$ are constant parameters. In Figure $\ref{3}$, we present the behaviour of two swampland parameters $c$ and $c'$ versus the width $\Gamma$ and the amplitude $\xi$ of the Lorentzian function when $|\beta|=0.02$ and $N=60$. From the panels (a) and (c), we find the swampland conditions $c\leq0.07$ and $c'\leq-0.012$ when the observationally favoured values $\Gamma=0.1$ and $\xi\leq600$ are considered. Also, the panels (b) and (d) almost present the same conditions when we focus on the values $\xi=100$ and $\Gamma\leq0.5$ which are compatible with the observations.

\section{Conclusion}
In the present manuscript, we have investigated a  quintessential inflationary model which is driven by a Lorentzian function in the presence of the constant-roll condition. We have calculated the slow-roll parameters and then the spectral inflationary parameters in order to present a complete study of the model. Comparing the obtained results with the Planck in combination with the BK14+BAO, we have found that the observationally allowed region of the width of the Lorentzian function is $0.3<\Gamma\leq0.5$ at the $68\%$ CL and $\Gamma\leq0.3$ at the $95\%$ CL when $|\beta|=0.02$ and the amplitude $\xi$ is $\sim120$. This result is obtained for $N=60$ while the model is almost ruled out by the observations for $N=50$. As a secondary analysis, we have acquired the acceptable range of the amplitude $\xi$ of the Lorentzian function when the width is fixed $\Gamma=0.1$, $|\beta|=0.02$ and $N=60$. We have figured out that the amplitudes $400<\xi\leq600$ and $\xi\leq400$ are in good agreement with the CMB observations at the $68\%$ and $95\%$ CL, respectively. Finally, we have found the swampland conditions $c\leq0.07$ and $c'\leq-0.012$ using the observational constraints on the width and the amplitude of the Lorentzian. The issue would be investigated in our future research, in particular, how works the reheating process in the model. Moreover, studying the non-Gaussianity properties of the model could be very interesting since the future observations of CMB will take into account the non-Gaussianity factor in the power spectrum.
\bibliographystyle{ieeetr}
\bibliographystyle{ieeetr}
\bibliography{biblo}

\begin{thebibliography}{10}

\bibitem{Guth}
A.~H. Guth, ``The inflationary universe: A possible solution to the horizon and
  flatness problems,'' {\em Phys. Rev. D}, vol.~23, p.~347, (1981).

\bibitem{Kazanas:1980tx}
D.~Kazanas, ``Dynamics of the universe and spontaneous symmetry breaking,''
  {\em Astrophys. J.}, vol.~241, p.~L59, (1980).

\bibitem{Linde:1981my}
A.~D. Linde, ``A new inflationary universe scenario: A possible solution of the
  horizon, flatness, homogeneity, isotropy problems,'' {\em Phys. Lett. B},
  vol.~108, p.~389, (1982).

\bibitem{Albrecht:1982wi}
A.~Albrecht and P.~J. Steinhardt, ``Cosmology for grand unified theories with
  radiatively induced symmetry breaking,'' {\em Phys. Rev. Lett.}, vol.~48,
  p.~1220, (1982).

\bibitem{Lyth:1998xn}
D.~H. Lyth and A.~Riotto, ``Particle physics models of inflation and the
  perturbation,'' {\em Phys. Rept.}, vol.~314, p.~1, (1999).

\bibitem{Kofman2}
L.~Kofman, A.~D. Linde, and A.~A. Starobinsky, ``Reheating after inflation,''
  {\em Phys. Rev. Lett.}, vol.~73, p.~3195, (1994).

\bibitem{Shtanov}
Y.~Shtanov, J.~H. Traschen, and R.~H. Brandenberger, ``Universe reheating after
  inflation,'' {\em Phys. Rev. D}, vol.~51, p.~5438, (1995).

\bibitem{staro}
A.~A. Starobinsky, ``A new type of isotropic cosmological models without
  singularity,'' {\em Phys. Lett. B}, vol.~91, p.~99, (1980).

\bibitem{barrow}
J.~D. Barrow and S.~Cotsakis, ``Inflation and the conformal structure of higher
  order gravity theories,'' {\em Phys. Lett. B}, vol.~214, p.~515, (1988).

\bibitem{kallosh1}
R.~Kallosh and A.~Linde, ``Universality class in conformal inflation,'' {\em
  JCAP}, vol.~07, p.~002, (2013).

\bibitem{martin}
J.~Martin, ``What have the planck data taught us about inflation?,'' {\em
  Class. Quant. Grav.}, vol.~33, p.~034001, (2016).

\bibitem{Chen}
X.~Chen, ``Primordial non-gaussianities from inflation models,'' {\em Adv.
  Astron.}, vol.~2010, p.~638979, (2010).

\bibitem{martin2}
J.~Martin, H.~Motohashi, and T.~Suyama, ``Primordial non-gaussianities from
  inflation models,'' {\em Phys. Rev. D}, vol.~87, p.~023514, (2013).

\bibitem{Motohashi1}
H.~Motohashi, A.~A. Starobinsky, and J.~Yokoyama, ``Inflation with a constant
  rate of roll,'' {\em JCAP}, vol.~09, p.~018, (2015).

\bibitem{Motohashi2}
H.~Motohashi and A.~A. Starobinsky, ``Constant-roll inflation: confrontation
  with recent observational data,'' {\em Europhys. Lett.}, vol.~117, p.~39001,
  (2017).

\bibitem{Inoue}
S.~Inoue and J.~Yokoyama, ``Curvature perturbation at the local extremum of the
  inflaton's potential,'' {\em Phys. Lett. B}, vol.~524, p.~15, (2002).

\bibitem{Pattison}
C.~Pattison, V.~Vennin, H.~Assadullah, and D.~Wandsa, ``The attractive
  behaviour of ultra-slow-roll inflation,'' {\em JCAP}, vol.~08, p.~048,
  (2018).

\bibitem{Kinney}
W.~H. Kinney, ``Horizon crossing and inflation with large $\eta$,'' {\em Phys.
  Rev. D}, vol.~72, p.~023515, (2005).

\bibitem{Namjoo}
M.~H. Namjoo, H.~Firouzjahi, and M.~Sasaki, ``Violation of non-gaussianity
  consistency relation in a single-field inflationary model,'' {\em Europhys.
  Lett.}, vol.~101, p.~39001, (2013).

\bibitem{Contaldi}
C.~R. Contaldi, L.~K. M.~Peloso, and A.~D. Linde, ``Suppressing the lower
  multipoles in the cmb anisotropies,'' {\em JCAP}, vol.~0307, p.~002, (2003).

\bibitem{Lello}
L.~Lello and D.~Boyanovsky, ``Tensor to scalar ratio and large scale power
  suppression from pre-slow roll initial conditions,'' {\em JCAP}, vol.~1405,
  p.~029, (2014).

\bibitem{Hazra}
D.~K. Hazra, A.~Shafieloo, G.~F. Smoot, , and A.~A. Starobinsky, ``Whipped
  inflation,'' {\em Phys. Rev. Lett.}, vol.~113, p.~071301, (2014).

\bibitem{Odintsov}
S.~Odintsov and V.~Oikonomou, ``Inflationary dynamics with a smooth slow-roll
  to constant-roll era transition,'' {\em Phys. Rev. D}, vol.~96, p.~024029,
  (2017).

\bibitem{Nojiri}
S.~Nojiri, D.~Odintsov, and V.~K. Oikonomou, ``Constant-roll inflation in
  \textit{f(R)} gravity,'' {\em Class. Quant. Grav.}, vol.~34, p.~245012,
  (2017).

\bibitem{Sebastiani}
S.~D. Odintsov, V.~K. Oikonomou, and L.~Sebastiani, ``Unification of
  constant-roll inflation and dark energy with logarithmic
  \textit{R}$^{2}$-corrected and exponential \textit{f(R)} gravity,'' {\em
  Nucl. Phys. B}, vol.~923, p.~608, (2017).

\bibitem{Motohashi9}
H.~Motohashi and A.~A. Starobinsky, ``\textit{f(R)} constant-roll inflation,''
  {\em Eur. Phys. J. C}, vol.~77, p.~538, (2017).

\bibitem{Cicciarella}
F.~Cicciarella, J.~Mabillard, and M.~Pieroni, ``New perspectives on
  constant-roll inflation,'' {\em JCAP}, vol.~01, p.~024, (2018).

\bibitem{Anguelova}
L.~Anguelova, P.~Suranyi, and L.~Wijewardhana, ``Systematics of constant roll
  inflation,'' {\em JCAP}, vol.~02, p.~004, (2018).

\bibitem{Ito}
A.~Ito and J.~Soda, ``Anisotropic constant-roll inflation,'' {\em Eur. Phys. J.
  C}, vol.~78, p.~55, (2018).

\bibitem{karam1}
A.~Karam, L.~Marzola, T.~Pappas, A.~Racioppi, and K.~Tamvakis, ``Constant-roll
  (quasi-)linear inflation,'' {\em JCAP}, vol.~05, p.~011, (2018).

\bibitem{Ghersi}
J.~T.~G. Ghersi, A.~Zucca, y, and A.~V. Frolov, ``Observational constraints on
  constant roll inflation,'' {\em JCAP}, vol.~05, p.~030, (2019).

\bibitem{Lin}
W.-C. Lin and M.~J.~P. Morsey, ``Dynamical analysis of attractor behavior in
  constant roll inflation,'' {\em JCAP}, vol.~09, p.~063, (2019).

\bibitem{Micu}
A.~Micu, ``Two-field constant roll inflation,'' {\em JCAP}, vol.~11, p.~003,
  (2019).

\bibitem{Oliveros}
A.~Oliveros and H.~E. Noriega, ``Constant-roll inflation driven by a scalar
  field with nonminimal derivative coupling,'' {\em Int. J. Mod. Phys. D},
  vol.~28, p.~1950159, (2019).

\bibitem{Motohashi3}
H.~Motohashi and A.~A. Starobinsky, ``Constant-roll inflation in scalar-tensor
  gravity,'' {\em JCAP}, vol.~11, p.~025, (2019).

\bibitem{oi1}
V.~Oikonomou, ``Reheating in constant-roll $f(r)$ gravity,'' {\em Mod. Phys.
  Lett. A}, vol.~32, p.~1750172, (2017).

\bibitem{neee}
H.~M. Sadjadi and V.~Anari, ``End of the constant-roll inflation, and the
  reheating temperature,'' {\em Phys. Dark Universe}, vol.~27, p.~100474,
  (2020).

\bibitem{oi2}
V.~K. Oikonomou and F.~P. Fronimos, ``A nearly massless graviton in
  einstein-gauss-bonnet inflation with linear coupling implies constant-roll
  for the scalar field,'' {\em EPL}, vol.~131, p.~30001, (2020).

\bibitem{Kamali}
V.~Kamali, M.~Artymowski, and M.~R. Setare, ``Constant roll warm inflation in
  high dissipative regime,'' {\em JCAP}, vol.~07, p.~002, (2020).

\bibitem{diego}
M.~Guerrero, D.~Rubiera-Garcia, and D.~S.-C. Gomez, ``Constant roll inflation
  in multifield models,'' {\em Phys. Rev. D}, vol.~102, p.~123528, (2020).

\bibitem{sh1}
M.~Shokri, J.~Sadeghi, M.~R. Setare, and S.~Capozziello, ``Nonminimal coupling
  inflation with constant slow roll,'' {\em To be publish in Int. J. Mod. Phys.
  D}, (2021).

\bibitem{sh2}
M.~Shokri, J.~Sadeghi, and M.~R. Setare, ``The generalized $sl(2, \textit{R})$
  and $su(1, 1)$ in non-minimal constant-roll inflation,'' {\em Ann. Phys.},
  vol.~429, p.~168487, (2021).

\bibitem{oki}
V.~K. Oikonomou, ``Generalizing the constant-roll condition in scalar
  inflation,'' {\em [gr-qc/2106.10778]}.

\bibitem{sh5}
M.~Shokri, J.~Sadeghi, and M.~R. Setare, ``Constant-roll inflation from a
  fermionic field,'' {\em [gr-qc/2107.03283]}.

\bibitem{shokriii}
M.~Shokri, M.~R. Setare, S.~Capozziello, and J.~Sadeghi, ``Constant-roll
  $f(\textit{R})$ inflation compared with cosmic microwave background
  anisotropies and swampland criteria,'' {\em [gr-qc:2108.00175]}.

\bibitem{vil}
P.~J.~E. Peebles and A.~Vilenkin, ``Quintessential inflation,'' {\em Phys. Rev.
  D}, vol.~59, p.~063505, (1999).

\bibitem{benn1}
D.~Benisty and E.~I. Guendelman, ``Lorentzian quintessential inflation,'' {\em
  Int. J. Mod. Phys. D}, vol.~29, p.~2042002, (2020).

\bibitem{ben1}
D.~Benisty and E.~I. Guendelman, ``Quintessential inflation from lorentzian
  slow roll,'' {\em Eur. Phys. J. C}, vol.~80, p.~577, (2020).

\bibitem{ben2}
L.~A. Saló, D.~Benisty, E.~I. Guendelman, and J.~d.~Haro, ``Quintessential
  inflation and cosmological seesaw mechanism: reheating and observational
  constraints,'' {\em JCAP}, vol.~07, p.~007, (2021).

\bibitem{benn2}
L.~A. Saló, D.~Benisty, E.~I. Guendelman, and J.~d.~Haro,
  ``$\alpha$-attractors in quintessential inflation motivated by
  supergravity,'' {\em Phys. Rev. D}, vol.~103, p.~123535, (2021).

\bibitem{cmb}
Y.~Akrami {\em et~al.}, ``Planck 2018 results. constraints on inflation,'' {\em
  Astron. Astrophys.}, vol.~641, p.~A10, (2020).

\bibitem{Vafa}
C.~Vafa, ``The string landscape and the swampland,'' {\em [hep-th/0509212]},
  (2005).

\bibitem{Brennan}
T.~D. Brennan, F.~Carta, and C.~Vafa, ``The string landscape, the swampland,
  and the missing corner,'' {\em PoS TASI}, vol.~77, pp.~015--072, (2017).

\bibitem{Ooguri}
H.~Ooguri and C.~Vafa, ``On the geometry of the string landscape and the
  swampland,'' {\em Nucl. Phys. B}, vol.~766, pp.~21--33, (2007).

\bibitem{Palti}
E.~Palti, ``The swampland: Introduction and review,'' {\em Fortsch. Phys.},
  vol.~67, pp.~1900037--33, (2019).

\bibitem{Palti2}
E.~Palti, ``Weak gravity conjecture and scalar fields,'' {\em JHEP}, vol.~1708,
  pp.~034--, (2017).

\bibitem{Obied}
G.~Obied, H.~Ooguri, L.~Spodyneiko, and C.~Vafa, ``De sitter space and the
  swampland,'' {\em [hep-th/1806.08362]}.

\end{thebibliography}
\appendix
\section{The slow-roll and spectral parameters of the model}
\vspace{0.8cm}
The slow-roll parameters of the model:
\begin{equation}
\epsilon=\frac{\pi^{2}(\beta-12)^{2}\cosh^{2}x+\Big(4\pi\beta\xi(\beta-12)\sinh x+(24\beta-288)\pi^{2}+4\beta^{2}\xi^{2}\Big)\cosh^{2}x+48\pi\beta\xi\sinh x-4\beta^{2}\xi^{2}+144\pi^{2}}{2\pi\Gamma\cosh^{2}x\sinh^{2}x\Big(36\pi^{2}\cosh^{2}x-12\pi\beta\xi\sinh x+\beta^{2}\xi^{2}-36\pi^{2}\Big)},
\label{a1}
\end{equation}
\begin{equation}
\eta=\frac{\pi^{2}(12-\beta)\cosh^{4}x-\Big(6\pi\xi(\beta-4)\sinh x+(\beta+24)\pi^{2}+4\beta\xi^{2}\Big)\cosh^{2}x+2\pi\xi(\beta-12)\sinh x+4\beta\xi^{2}+12\pi^{2}}{\cosh^2x\Big(6\pi^{2}\Gamma\sinh^{3}x-\Gamma\pi\xi\beta\sinh^{2}x\Big)},
\label{a2}
\end{equation}
\begin{eqnarray}
&\!&\!\zeta^{2}=\frac{1}{{\pi^{2}\Gamma^{2}\cosh^{4}x\sinh^{4}x\Big(36\pi^{2}\cosh^{2}x-12\pi\beta\xi\sinh x+\beta^{2}\xi^{2}-36\pi^{2}\Big)}}\Bigg\{\pi^{2}(\beta-12)^{2}\cosh^{8}x+\pi^{2}\cosh^{6}x\Big(16(\beta-\frac{9}{2})\times\nonumber\\&\!&\!
\times(\beta-12)\pi\xi\sinh x+5\pi^{2}(\beta-144)+4\xi^{2}(13\beta^{2}-120\beta+144)\Big)+\cosh^{4}x\Big(4\pi\xi\sinh x\big((\beta^{2}+120\beta-648)\pi^{2}+2\beta\xi^{2}\times\nonumber\\&\!&\!
\times(7\beta-24)\big)+1296\pi^{4}-4(19\beta^{2}-312\beta+432)\xi^{2}\pi^{2}+16\beta^{2}\xi^{4}\Big)+\cosh^{2}x\Big(4\pi\xi\sinh x\big((\beta^{2}-78\beta+648\big)\pi^{2}-4\beta(5\beta-\nonumber\\&\!&\!
-24)\xi^{2})+(24\beta-1008)\pi^{4}+32\xi^{2}(\beta^{2}-32\beta+54)\pi^{2}-32\beta^{2}\xi^{4}\Big)+24\xi\pi\sinh x\Big(4(\beta-9)\pi^{2}+\xi^{2}\beta(\beta-8)\Big)+288\pi^{4}-\nonumber\\&\!&\!
-8\xi^{2}(\beta^{2}-36\beta+72)\pi^{2}+16\beta^{2}\xi^{4}\Bigg\},
\label{a3}
\end{eqnarray}
where $x\equiv\sqrt{\frac{\pi}{\xi\Gamma}}\varphi.$\\\\
\vspace{0.5cm}
The spectral index of the mode:
\begin{eqnarray}
&\!&\!n_{s}=\frac{1}{\pi\big((\beta-12)^{2}N^{2}+9\Gamma^{2}\big)^{2}\big((\beta-12)N+3\Gamma\big)^{2}\big((\beta-12)N-3\Gamma\big)^{2}\Big(\pi(\beta-12)^{2}N^{2}-\xi\beta\Gamma N(\beta-12)-9\pi\Gamma^{2}\Big)^{2}}\nonumber\\&\!&\!
\times\Bigg\{\pi^{3}(\beta-12)^{12}N^{12}-2(\beta-12)^{11}\big((\Gamma\xi+\pi\big)\beta-12\pi)\pi^{2}N^{11}-\Big(\xi(-\Gamma\xi+\pi)\beta^{2}+24\pi\xi\beta+18\pi(\pi\Gamma+8\xi)\Big)(\beta-12)^{10}\nonumber\\&\!&\!
\times\Gamma\pi N^{10}-18(\beta-12)^{9}\Gamma^{2}\Big((-\pi\xi\Gamma+7\pi^{2}-16\xi^{2})\beta+60\pi^{2}\Big)\pi N^{9}-9\Big((-4\pi^{2}\xi+16\xi^{3})\beta^{2}-192\pi^{2}\xi\beta+9\pi^{2}(\pi\Gamma-64\xi)\Big)\nonumber\\&\!&\!
\times(\beta-12)^{8}\Gamma^{3}N^{8}-108(\beta-12)^{7}\Gamma^{4}\Big(8\beta^{2}\xi^{2}+(-3\pi\xi\Gamma+9\pi^{2}+72\xi^{2})\beta-180\pi^{2}\Big)\pi N^{7}+162(\beta-12)^{6}\Gamma^{5}\Big((-\pi\xi^{2}\Gamma+5\pi^{2}\xi\nonumber\\&\!&\!
+16\xi^{3})\beta^{2}-168\pi^{2}\xi\beta+18\pi^{2}(\pi\Gamma-24\xi)\Big)N^{6}+972(\beta-12)^{5}\Gamma^{6}\Big(8\beta^{2}\xi^{2}+(-3\pi\xi\Gamma+9\pi^{2}+72\xi^{2})\beta-180\pi^{2}\Big)\pi N^{5}-729\nonumber\\&\!&\!
\times\Big((-4\pi^{2}\xi+16\xi^{3})\beta^{2}-192\pi^{2}\xi\beta+9\pi^{2}(\pi\Gamma-64\xi)\Big)(\beta-12)^{4}\Gamma^{7}N^{4}+13122(\beta-12)^{3}\Gamma^{8}\Big((-\pi\xi\Gamma+7\pi^{2}-16\xi^{2})\beta+60\pi^{2}\Big)\nonumber\\&\!&\!
\times\pi N^{3}-6561\Big(\xi(-\Gamma\xi+\pi)\beta^{2}+24\pi\xi\beta+18\pi(\pi\Gamma+8\xi)\Big)(\beta-12)^{2}\Gamma^{9}\pi N^{2}+118098(\beta-12)\Gamma^{10}\big((\Gamma\xi+\pi)\beta-12\pi\big)\pi^{2}N\nonumber\\&\!&\!+531441\pi^{3}\Gamma^{12}\Bigg\}
\label{a4}
\end{eqnarray}
\vspace{0.5cm}
The tensor-to-scalar ratio of the model:
\begin{equation}
r=\frac{8N^{2}(\beta-12)^{4}\Gamma\xi\Big(\pi(\beta-12)^{4}N^{4}+12\xi\beta\Gamma(\beta-12)^{2}N^{3}+\big(18\pi\beta^{2}\Gamma^{2}-2592\pi\Gamma^{2}\big)N^{2}-108N\beta\Gamma^{3}\xi+81\pi\Gamma^{4}\Big)^{2}}{\pi\big((\beta-12)^{2}N^{2}+9\Gamma^{2}\big)^{2}\big((\beta-12)N+3\Gamma\big)^{2}\big((\beta-12)N-3\Gamma\big)^{2}\Big(\pi(\beta-12)^{2}N^{2}-\xi\beta\Gamma N(\beta-12)-9\pi\Gamma^{2}\Big)^{2}}
\label{a5}
\end{equation}

\end{document}